\documentclass[preprint]{elsarticle}
\usepackage{amssymb}
\usepackage{morefloats}
\usepackage{subfigure}

\journal{Astroparticle Physics}
\begin{document}
\begin{frontmatter}
  
  \title{The Energy Spectrum of Telescope Array's Middle
    Drum Detector and the Direct Comparison to the High Resolution Fly's Eye Experiment}
  
   \author[label1]{T.~Abu-Zayyad}
   \author[label2]{R.~Aida}
   \author[label1]{M.~Allen}
   \author[label1]{R.~Anderson}
   \author[label3]{R.~Azuma}
   \author[label1]{E.~Barcikowski}
   \author[label1]{J.~W.~Belz}
   \author[label1]{D.~R.~Bergman}
   \author[label1]{S.~A.~Blake}
   \author[label1]{R.~Cady}
% ...

   \address[label1]{University of Utah, High Energy Astrophysics Institute, Salt Lake City, Utah, USA}
   \address[label2]{University of Yamanashi, Interdisciplinary Graduate School of Medicine and Engineering, Kofu, Yamanashi, Japan}
   \address[label3]{Tokyo Institute of Technology, Meguro, Tokyo, Japan}
% ...

  \cortext[cor1]{Corresponding author}
  \fntext[email]{doug@cosmic.utah.edu}
 
  \begin{abstract}
    The Telescope Array's Middle Drum fluorescence detector was
    instrumented with
    telescopes refurbished from the High Resolution Fly's Eye's
    HiRes-1 site. The data observed by Middle Drum in monocular mode
    was analyzed via the HiRes-1 profile-constrained geometry
    reconstruction technique and utilized the same calibration
    techniques enabling a direct comparison of the
    energy spectra and energy scales between the two experiments. The 
    spectrum measured using the Middle Drum telescopes is based on a three-year exposure collected between
    December 16, 2007 and December 16, 2010. The 
    calculated difference between the spectrum of the Middle Drum observations
    and the published spectrum obtained by the data collected by the
    HiRes-1 site allows the HiRes-1 energy scale to be transferred to
    Middle Drum. The HiRes energy scale is applied to
    the entire Telescope Array by making a comparison between Middle
    Drum monocular events and hybrid events that triggered both Middle
    Drum and the Telescope Array's scintillator Ground Array.
  \end{abstract}


\end{frontmatter}

\section{Telescope Array}
\label{TAMD}

\begin{figure}[tp]
  \centerline{
    \includegraphics[width=0.6\textwidth]{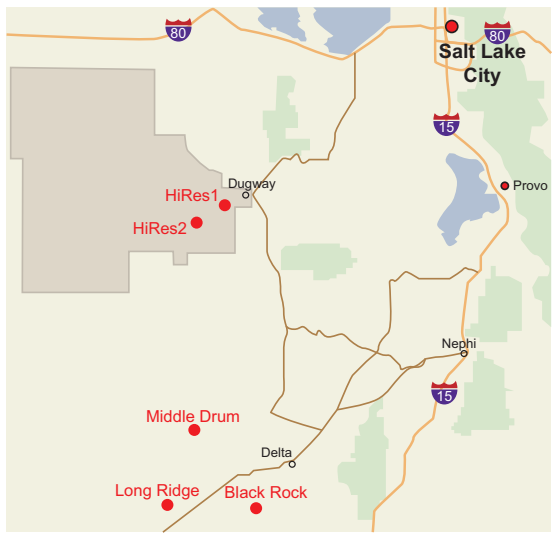}
  }
  \caption{Map showing the location of Telescope Array relative to
    Salt Lake City and Dugway, Utah (the location of the High
    Resolution Fly's Eye). The route from Salt Lake City to Delta is
    136 miles.}
  \label{fig:SLC_to_MD}
\end{figure}

\begin{figure}[bp]
  \centerline{
   \includegraphics[width=\textwidth]{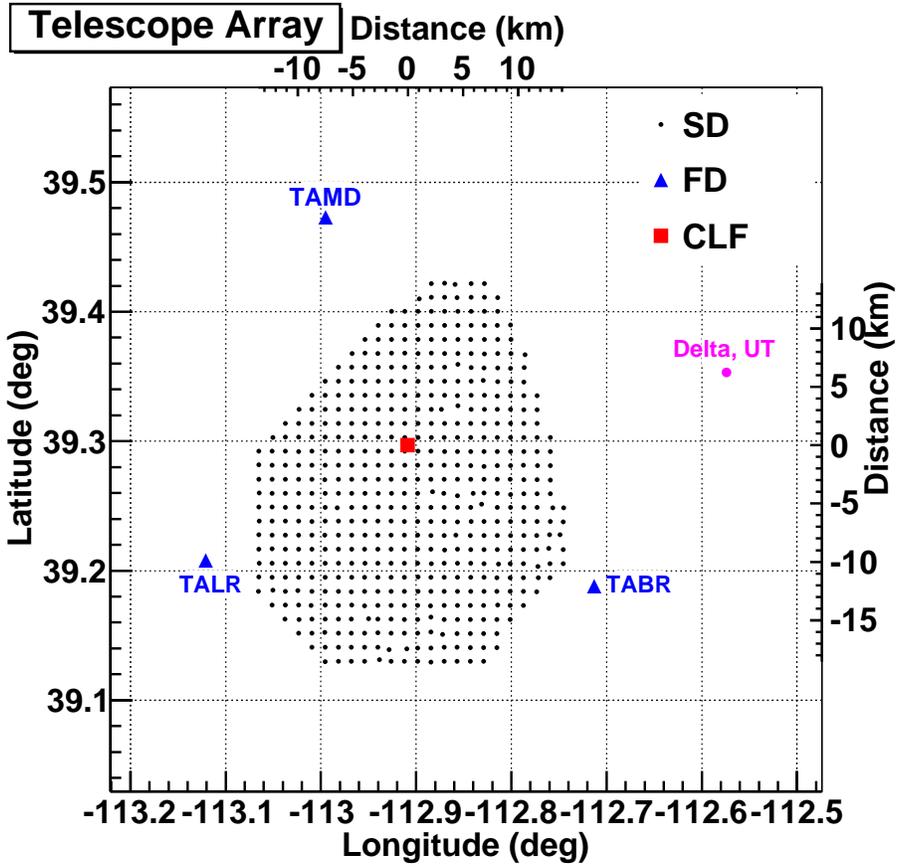}
  }
  \caption{Map of the Telescope Array detectors. The telescope
    stations (Middle Drum, Long Ridge, and
    Black Rock Mesa) are indicated by blue triangles. The scintillator
    detectors are indicated by black dots and the Central Laser
    Facility by a red square. The ground array occupies about 700
    $km^{2}$ west of Delta, Utah.}
  \label{fig:TA_Map}
\end{figure}

The Telescope Array (TA) is the largest cosmic ray experiment in the
northern hemisphere. It was designed to help
resolve physics differences between the High Resolution Fly's Eye
(HiRes) experiment, the Akeno Giant Air
Shower Array (AGASA) \cite{AGASA}, and the Pierre Auger Observatory
\cite{PAO}. TA consists of three HiRes-like fluorescence telescope
stations overlooking 507 AGASA-like scintillator surface detectors (see Figures
\ref{fig:SLC_to_MD} and \ref{fig:TA_Map}). The Surface Detector (SD)
array was deployed in a square grid with a 1.2 km separation,
covering $\sim700~\rm{km^{2}}$ \cite{TASD_ICRC}. Each SD unit is
composed of two layers of $3~\rm{m^{2}} \times 1.2~\rm{cm}$ scintillating
plastic sheets separated by a thin steel sheet. The light from each
layer is collected by wavelength-shifting optical fibers and directed
into separate Photo-Multiplier Tubes (PMTs). 


Three telescope stations view the sky over the scintillator array. The northernmost fluorescence station, known as the Middle Drum site, consists
of 14 telescopes refurbished from the HiRes experiment's HiRes-1
site. These were deployed between November, 2006 and October, 2007 and
were arranged to view $\sim 120^{\circ}$ in azimuthal and
$3^{\circ} - 31^{\circ}$ in elevation. Compared to HiRes-1
\cite{HiRes-1_Spect}, the Middle Drum site has only $1/3$ of the
azimuthal coverage but observes twice the elevation, as it
was deployed into two rings, each covering $14^{\circ}$ in elevation. Each
telescope unit uses sample-and-hold electronics with a $5.6~\mu s$
gate. Each telescope camera consists of 256 PMTs covered with an ultra-violet
band-pass filter. Descriptions of the Black Rock and Long Ridge telescope stations
were described by Tokuno \cite{TAFD_Paper}.

The goals of the Middle Drum spectral analysis are
three-fold. The primary goal of this analysis is to determine the flux
of particles using the same calibration and analysis processing tools
used to produce the monocular spectrum from the HiRes-1 data. The second goal is to compare
the spectra measured by the Middle Drum detector with that of HiRes-1. Since the telescope
units used in both of these detectors are composed
of the same equipment, the results of this comparison produce a direct link in
the energy scale between these two experiments. Finally, by comparing
events observed by Middle Drum and any of the other TA detectors, the
energy scale of the entire Telescope Array experiment can be
compared to that of the HiRes experiment. In this paper, this
comparison is performed between the geometries of the events observed
by Middle Drum and reconstructed using the monocular technique to those events that triggered both Middle
Drum and the SD array and analyzed using a hybrid technique.

\section{Event Reconstruction and Selection}
\label{Anal}

The Middle Drum data and Monte Carlo events (described in section
\ref{MC}) were processed using the same
programs created for HiRes-1 analysis \cite{DCR_thesis}. The only
changes made were for the location and pointing directions of the
telescopes. The HiRes-1 analysis was unique in that there was 
limited elevation coverage and a traditional monocular reconstruction
could not be performed on the data. Instead, a combined
geometrical-profile reconstruction was developed by Abu-Zayyad
\cite{tareq_thesis} which increased the resolution of the observed
showers. This technique was not required in the analysis of Middle
Drum data since the detector observed longer-track events due to the
increase in elevation angle coverage, however, it was
used for consistency.

As at HiRes, lasers are used for light-attenuation calibration,
aerosol measurements, and relative-timing variances between the three
fluorescence detector sites. Most of the events observed by
the Middle Drum detector belong to these calibration lasers which are primarily removed by 
only processing those events that are downward-tending, since the lasers
are fired in either upward or horizontal directions. Some of these
laser shots appear to be downward-going events due to preliminary
calculations using the timing and pointing directions of the triggered tubes. These are removed using the GPS
trigger time-stamp and the GPS measured site positions.

After filtering out laser events, most of those events that remain are due to electronic noise
triggers, airplanes, and muons that pass through the camera's PMTs. These are removed by determining a
correlation between the time and geometrical pattern of the triggered
tubes. Triggered tubes are clustered into groups of three or more
tubes with difference limits on the trigger-time of $2.0~\mu s$ and
the pointing-direction of $1.2^{\circ}$
from the previously triggered tube. These clusters are then combined
into a single event-track from which a shower-detector plane (SDP) is
determined. The tubes in a track are then iteratively checked and
removed if greater than 3 RMS deviations away in either time or angle
from the mean \cite{DCR_thesis}. 

The Middle Drum data and Monte Carlo simulations are 
reconstructed in monocular mode with the geometry determined by the
equation 
\begin{equation}
  t_{i} = t_{0} + \frac{ R_{p} }{ c }\tan \left( \frac{ \pi - \psi - \chi_{i} }{ 2 } \right),
  \label{eq:time_fit}
\end{equation}
\noindent where $t_{i}$ and $\chi_{i}$ are the respective trigger time
and pointing direction of tube $i$, $R_{P}$ is the impact parameter of
the shower with respect to the detector, $\psi$ is the angle the axis
of the shower makes with respect to the direction of the core impact
position around the detector, and $t_{0}$ is the time the shower is
calculated to be at $R_{P}$. 

The profile of the shower is calculated using the Gaisser-Hillas
parameterization \cite{GHEquation}
\begin{equation}
  N_{e}(x) = N_{max} \times \left[ \frac{x - X_{0}}{X_{max} - X_{0}} \right] ^{\frac{X_{max} - X_{0}}{\lambda}} e^{\frac{X_{max} - x}{\lambda}},
  \label{eq:Gaisser-Hillas_Eq}
\end{equation}
\noindent where $N_{e}(x)$ is the number of charged particles
(measured from the signal strength) at a given slant depth, $x$, in
$\rm{g/cm^{2}}$; $N_{max}$ is the maximum number of secondary
particles produced in the extensive air shower, located at $X_{max}$;
$X_{0}$ is a fit parameter associated with the depth of the first
interaction; and $\lambda$ is a fit paramater defining the width of
the shower profile. 

To reconstruct the Middle Drum data, as was done for HiRes-1, the
time-fit was constrained by the shower profile reconstruction. This
was performed by setting $\lambda$ to a constant $60~\rm{g/cm^{2}}$
and $X_{0}$ to a constant $-100~\rm{g/cm^{2}}$ in
order to constrain the width and initial depth of the shower. These constants are 
in good agreement with average simulated shower measurements \cite{DCR_thesis}. An inverse-Monte Carlo
reconstruction is then made by 
simulating showers that closely resemble the true event using the triggered tubes. This is
performed by choosing a series of $X_{max}$ values for individual
Monte Carlo events over all energies in the shower library.  To determine a best-fit profile reconstruction, a
comparison is made between the light signal actually observed to the
one simulated for each tube considered in the reconstruction. This is
effective since both the timing and the profile fits are only dependent on
the trigger time and pointing directions of the tubes used in the
reconstruction, which determine the slant depth of the
shower that each tube is observing along the axis of the shower. 

Separate chi-square minimizations are then performed on the timing and
the profile reconstructions for each of the constant $X_{max}$
values chosen. The timing chi-square is calculated by
\begin{equation}
  \chi_{time}^{2} = \sum_{i} \frac{ 1 }{ \sigma_{i}^{2} } \left\{ t_{i} - \left[ t_{0} + \left( \frac{ R_{p} }{ c } \right) \tan \left( \frac{ \pi - \psi - \chi_{i} }{ 2 } \right) \right] \right\} ^{2}
  \label{eq:Time_Chi2}
\end{equation}
\noindent with the error, $\sigma_{i}$, determined by the time to cross
the face of a PMT. The
profile chi-square is calculated by 
\begin{equation}
  \chi_{profile}^{2} = \sum_{i} \left( \frac{1}{\sigma_{i}^{2}} \right) (S_{i}^{o} - S_{i}^{e})^{2}
  \label{eq:Prof_Chi2}
\end{equation}
\noindent where, as in the timing fit, the sum is performed over the
tubes within 3 RMS deviations away from the shower-detector plane,
The observed signal, $S_{i}^{o}$, is also used to calculated the
uncertainty, $\sigma_{i}^{2}$, which
is estimated to be $S_{i}^{o} + S$. The constant, $S = 200$, is obtained through adding
in quadrature the sky noise  and electronic fluctuations. Details of
the reconstruction codes can be found in the dissertation by Abu-Zayyad
\cite{tareq_thesis}. 

The optimal reconstruction is then determined by
calculating a best combined chi-squared for each $X_{max}$ fit using 
\begin{equation}
  \chi_{comb}^{2} = \overline{\chi}_{profile}^{2} + \overline{\chi}_{time}^{2}
  \label{eq:IMINC}
\end{equation}
\noindent where $\overline{\chi}$ is the normalized chi-square value calculated as
\begin{equation}
  \overline{\chi} = \chi_{fit} \times \frac{NDF_{min}}{\chi_{min}}
\end{equation}
\noindent where $\chi_{fit}$ is the chi-square of from the fit for
each $X_{max}$ and $NDF_{min}$ and $\chi_{min}$ are the number of
degrees of freedom and the chi-square, respectively, for the smallest
chi-square reconstruction. As mentioned previously, this innovative technique was developed to
reconstruct HiRes-1 data which had a limited elevation
coverage. Future analyses of Middle
Drum data will include traditional monocular reconstruction
techniques. However, this method was used in this current analysis to
provide a direct comparison to the spectrum observed by
HiRes-1. Additionally, this technique results in a better
resolution and aperture than an unconstrained time fit, even for the
longer tracks observed at the Middle Drum site. 

After the selection of candidate events is obtained, quality
cuts are performed on the fully reconstructed showers to remove any
event that exhibits anomalous behavior. These cuts were optimized for the the short
shower tracks observed by HiRes-1 \cite{tareq_thesis}. The HiRes-1 analysis was ideal for
cosmic rays with energy greater than $10^{18} eV$ since they could be
observed from farther away and would appear as short tracks in the
lower elevations. These cuts are applied to Middle Drum events since the
higher-energy events would still have shorter tracks and the
overlapping energy range between HiRes-1 and Middle Drum could then be
directly compared. Additionally, this gives a baseline to future analyses. Events are retained if: 
\begin{enumerate}
\item the event reconstruct well, as determined by
  \begin{itemize}
    \item not rejecting too many off-plane tubes,
    \item there are enough slant-depth bins to fit a profile,
    \item a $\chi^{2}$ minimum is attained, and
    \item the modified geometry still parameterizes the timing fit;
 \end{itemize}
\item the angular tracks are $\geq 7.9^{\circ}$, so that there are
  enough triggered tubes to provide a reliable reconstruction;
\item the shower depth into the atmosphere observed by the first tube used
  in the reconstruction is $< 1000~\rm{g/cm^{2}}$, so the fit is not
  focusing on the tail of the shower;
\item the in-plane angle, $\psi$, is $< 120^{\circ}$, to make sure the
  detector is not overwhelmed with \v{C}erenkov radiation; and
\item the area of the mirror observing the fitted track (away from the
  mirror/tube edges) is $> 0.9~\rm{m^{2}}$, to ensure there is not a
  bias in the reconstructed signal strength.
\end{enumerate}

\section{Monte Carlo Simulation}
\label{MC}

\begin{figure}[htp]
  \centerline{
    \includegraphics[width=\textwidth]{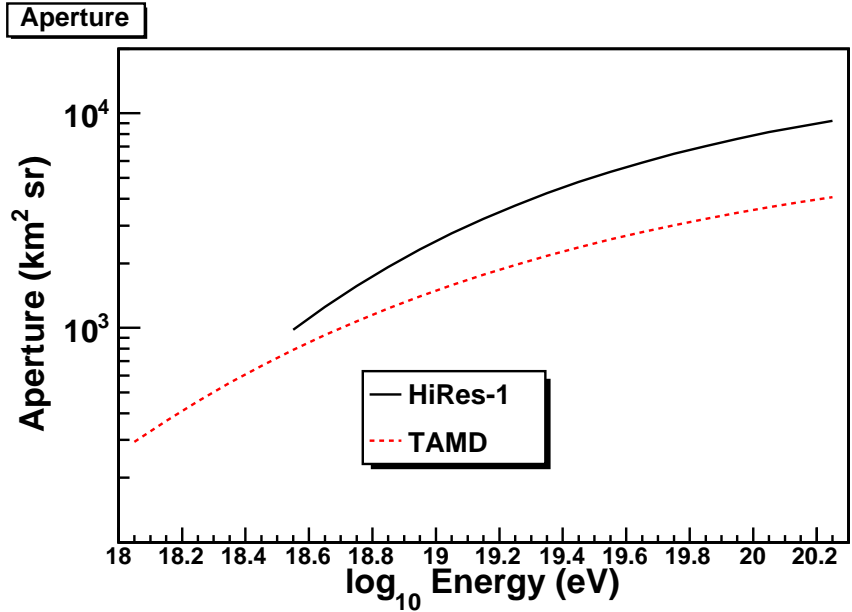}
  }
  \caption{The calculated Middle Drum aperture compared to that for the HiRes-1 detector.} 
  \label{fig:TAMD_HR1_Aper}
\end{figure}

The energy-dependent aperture of the detector is the product of the effective area and the
solid angle of acceptance. This is calculated using the equations 
\begin{equation}
  (A\Omega)_{0} (\rm{m^{2} ster}) = 2 \pi^{2} \left( R^{2}_{p~max} - R^{2}_{p~min} \right) \times \left( 1 - \cos\theta_{max} \right)
  \label{eq:Aper_Zero}
\end{equation}
\noindent and
\begin{equation}
  A \Omega (E) = (A \Omega)_{0} \frac{N_{recon}(E_{MC-recon})}{N_{gen}(E_{MC-gen})},
  \label{eq:Aperture_calc}
\end{equation}
\noindent where $R_{P}$ is the distance of closest approach of the
shower, $\theta$ is the zenith angle of the shower, $N_{recon}$ is the
number of events reconstructed with energy,
$E_{MC-recon}$, and $N_{gen}$ is the number of events generated with
energy $E_{MC-gen}$. Counting the number of events reconstructed at an
energy folds in the detector bias into the energy spectrum
calculation. Alternatively, the detector efficiency at a certain energy can be calculated by replacing
$N_{recon}(E_{MC-recon})$ with the number of events retained with a certain
generated energy, $N_{recon}(E_{MC-gen})$. The aperture of the Middle Drum detector has been
calculated to be $\sim60\%$ that of HiRes-1 for reconstructed energies of
$10^{19.0}$ eV, falling linearly to 40\% at $10^{20.0}$ eV (see Figure
\ref{fig:TAMD_HR1_Aper}). 

The CORSIKA-simulated shower library used by Middle Drum was the same
generated for HiRes, using QGSJET01 as the hadron interaction model \cite{HiRes-1_Spect}.
These showers were thrown with an isotropic distribution and consisted
of $\sim10\times$ the exposure of the Middle Drum collected data.
The Monte Carlo simulated only proton events between
$10^{17.5}$ eV and $10^{21.0}$ eV using values as measured by HiRes
below the GZK cutoff \cite{HiRes_GZK_paper} \cite{HiRes_Comp}. A 
spectral index of 3.25 was used below $10^{18.65}$ eV and 2.81 above. The spectral set was thrown without simulating the GZK
suppression \cite{PhysRevLett.16.748} \cite{SovPhysJETPL.4.78}. The
lower energy range was thrown out to a range of 25 km from the
telescope site, well beyond where the detector is incapable of
triggering on the fluorescence light of a $10^{18.65}$ eV cosmic ray
shower. The higher energy range was thrown out to 50
km. The simulated showers of both energy regions were thrown with a
maximum zenith angle of $80^{\circ}$. The CORSIKA output is fed into
the detector Monte Caro resulting in events which look exactly like
real data and are subjected to the same reconstruction programs and
quality cuts.

\subsection{Data-Monte Carlo Comparison}
\label{DTMC}

\begin{figure}[htp]
  \centering
  \includegraphics[width=\textwidth]{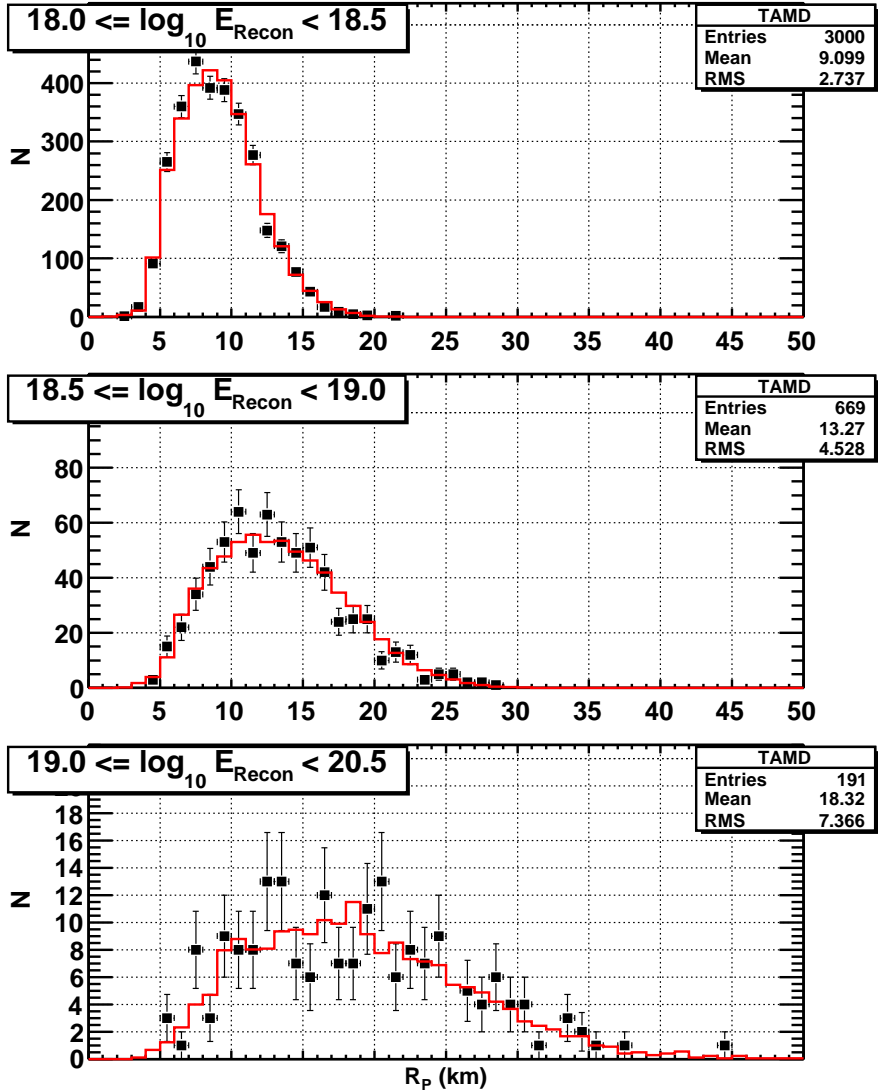}
  \caption{Comparison of the data and Monte Carlo distributions of the
    impact parameter, $R_{P}$, in three energy ranges: $10^{18.0-18.5}$ eV, $10^{18.5-19.0}$
eV, and $> 10^{19.0}$ eV. The Monte Carlo
    (red histogram) is in excellent agreement with the data (black
    points with error bars). The number of entries indicates the
    number of data events observed in that energy range.} 
  \label{fig:DTMC_Rp}
\end{figure}

\begin{figure}[htp]
  \centering
  \includegraphics[width=\textwidth]{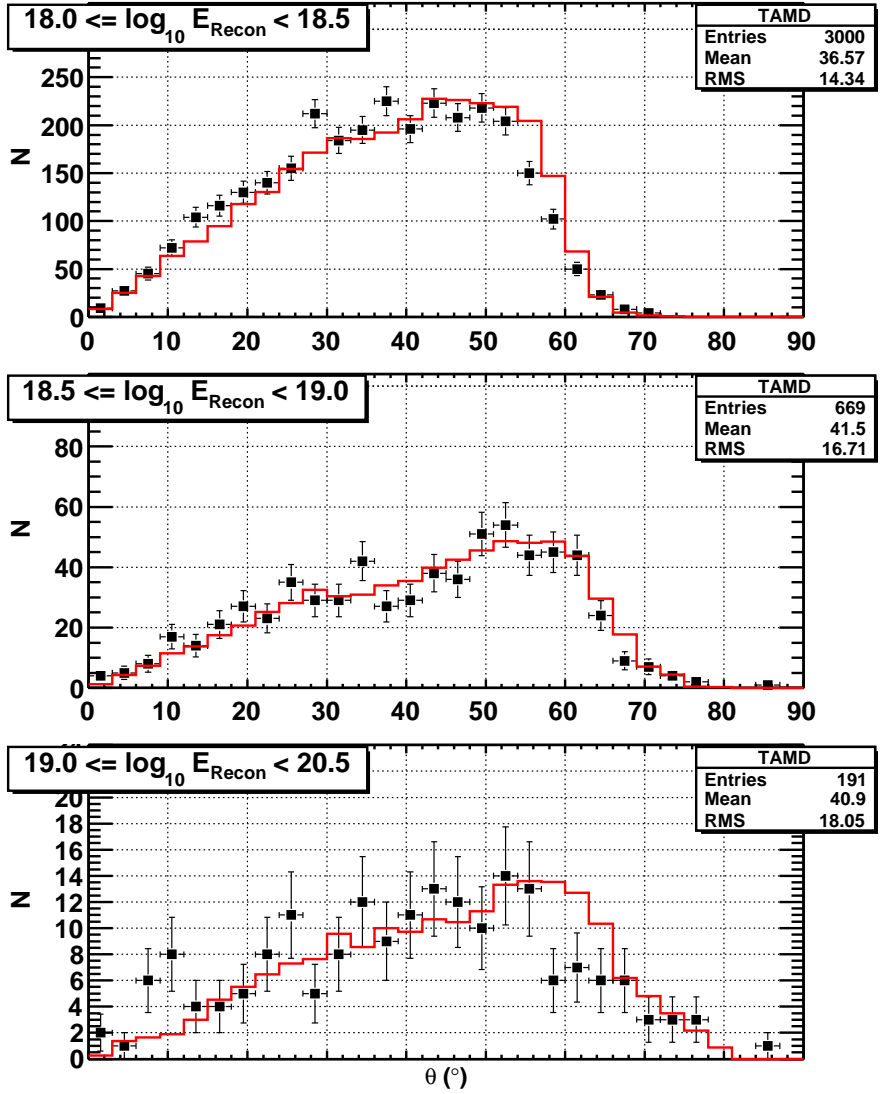}
  \caption{Comparison of the data and Monte Carlo distributions of the
    zenith angle, $\theta$, in three energy ranges. The Monte Carlo
    (red histogram) is in excellent agreement with the data (black
    points with error bars). The number of entries indicates the
    number of data events observed in that energy range.} 
  \label{fig:DTMC_Zen}
\end{figure}

\begin{figure}[htp]
  \centering
  \includegraphics[width=\textwidth]{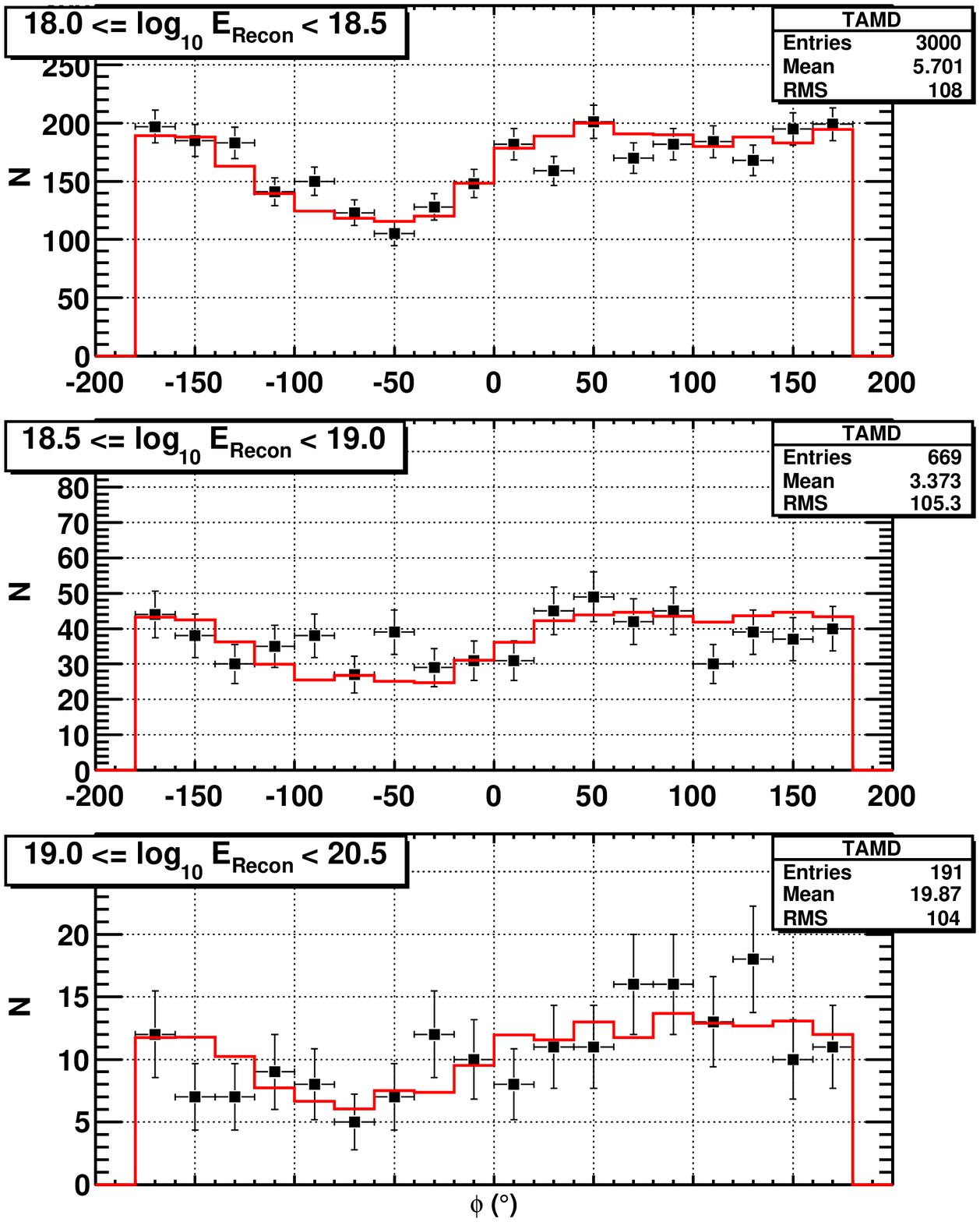}
  \caption{Comparison of the data and Monte Carlo distributions of the
    azimuthal angle, $\phi$, in three energy ranges: $10^{18.0-18.5}$
    eV, $10^{18.5-19.0}$ eV, and $> 10^{19.0}$ eV. The Monte Carlo
    (red histogram) is in excellent agreement with the data (black
    points with error bars). The number of entries indicates the
    number of data events observed in that energy range. The mean of
    this distribution indicates that most events are pointing away from the detector.} 
  \label{fig:DTMC_Azi}
\end{figure}

To verify the adequacy of the Monte Carlo used for the aperture
calculation and to confirm that the Monte Carlo
closely models the real data, it is important to compare the distributions of the reconstructed
Monte Carlo events and the data. These are shown in three
energy ranges ($10^{18.0-18.5}$ eV, $10^{18.5-19.0}$
eV, and $> 10^{19.0}$ eV) in order to demonstrate that the Monte Carlo has the same
energy evolution as the data. The variables chosen for this comparison are
those that directly determine the aperture: the impact parameter,
$R_{P}$ (Figure \ref{fig:DTMC_Rp}); the shower zenith angle,
$\theta$ (Figure \ref{fig:DTMC_Zen}); and the shower azimuthal
angle, $\phi$ (Figure \ref{fig:DTMC_Azi}). 

The impact parameter distribution directly determines the effective area of the aperture, and, as
expected, the mean of the distribution increases along with the
spread, or RMS, as the energy increases. The zenith and azimuthal angles
directly determine the solid angle of acceptance. For all three
parameters, the (black) data points and (red) Monte Carlo histogram
distributions are in excellent agreement. 

It should also be noted that since the Middle Drum telescopes are pointing
in the South-East direction, and that there is a quality cut removing
many of those events that are pointing towards the detector, there is
a depletion observed in the azimuthal distribution in this direction. This variance decreases with increasing energy since the
impact parameter moves farther away from the detector and, therefore, there are
fewer showers pointing above the $120^{\circ}$ limitation.

\subsection{Resolution}
\label{Resolution}

\begin{figure}[htp]
  \centering
  \includegraphics[width=\textwidth]{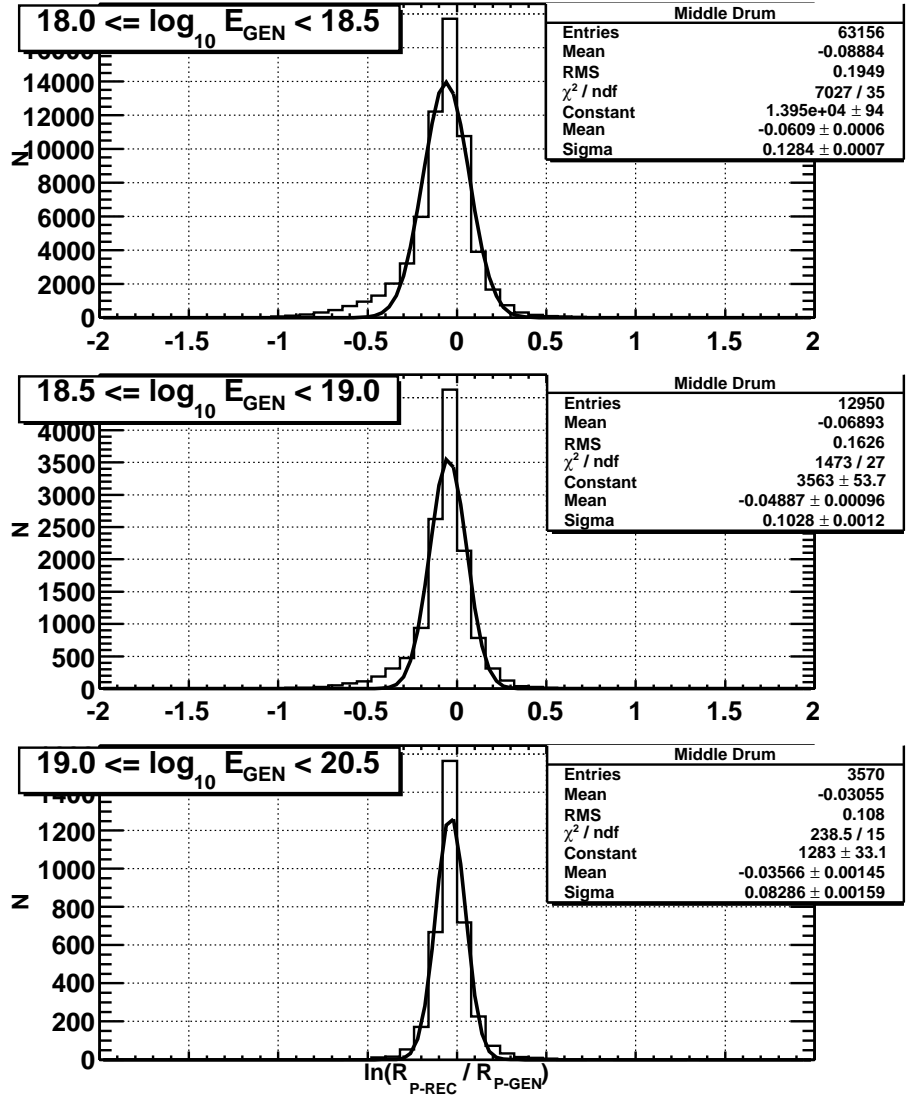}
  \caption{Resolution in the measurement of the impact parameter,
    $R_{P}$, of Monte Carlo simulated events shown in three energy
    ranges: $10^{18.0-18.5}$ eV, $10^{18.5-19.0}$ eV, and $> 10^{19.0}$ eV. The gaussian fit is used to determine the detector bias and resolution.} 
  \label{fig:Res_Rp}
\end{figure}

\begin{figure}[htp]
  \centering
  \includegraphics[width=\textwidth]{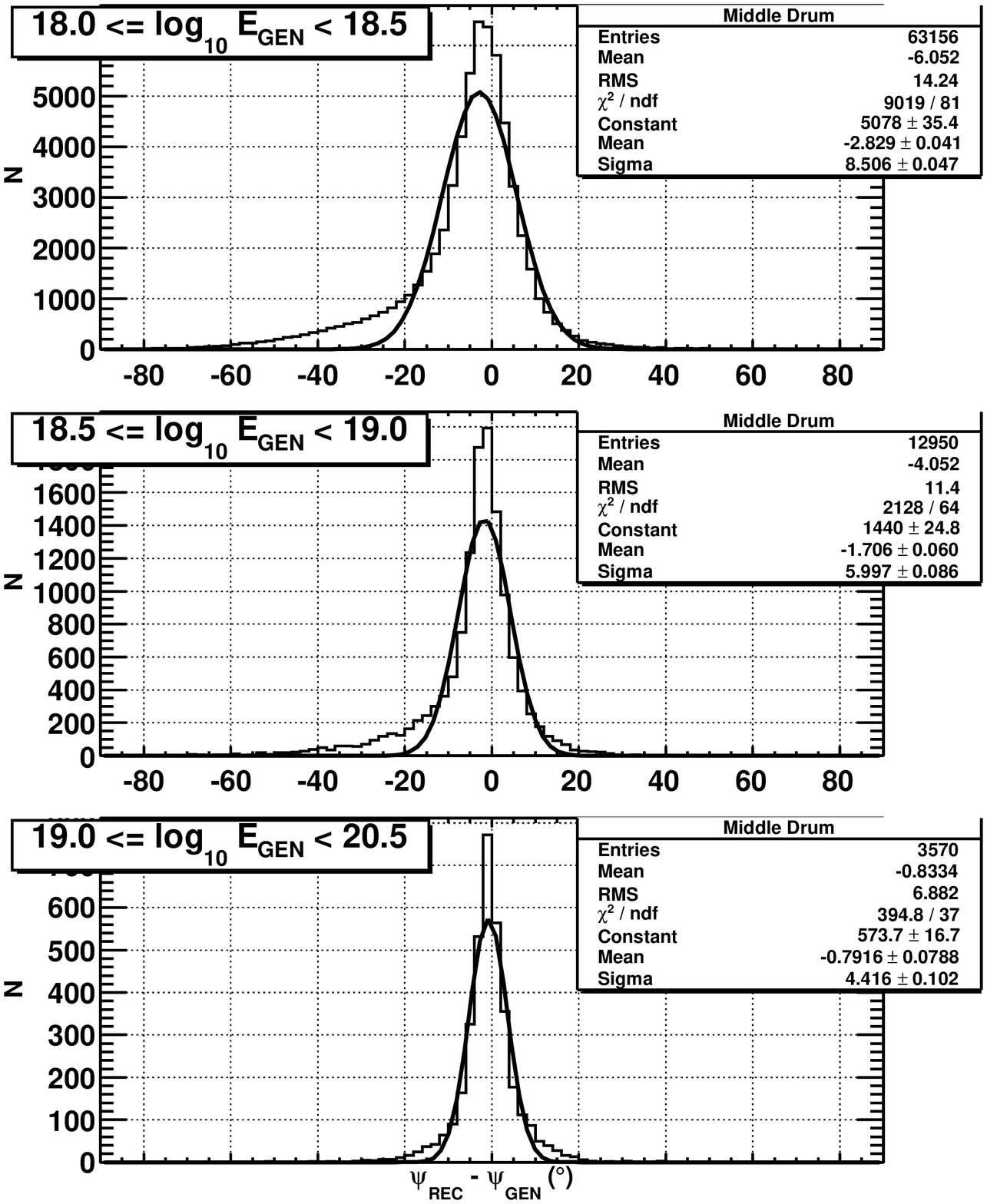}
  \caption{Resolution in the measurement of the in-plane angle,
    $\psi$, of Monte Carlo simulated events shown in three energy
    ranges: $10^{18.0-18.5}$ eV, $10^{18.5-19.0}$ eV, and $> 10^{19.0}$ eV. The gaussian fit is used to determine the detector bias and resolution.} 
  \label{fig:Res_Psi}
\end{figure}

\begin{figure}[htp]
  \centering
  \includegraphics[width=\textwidth]{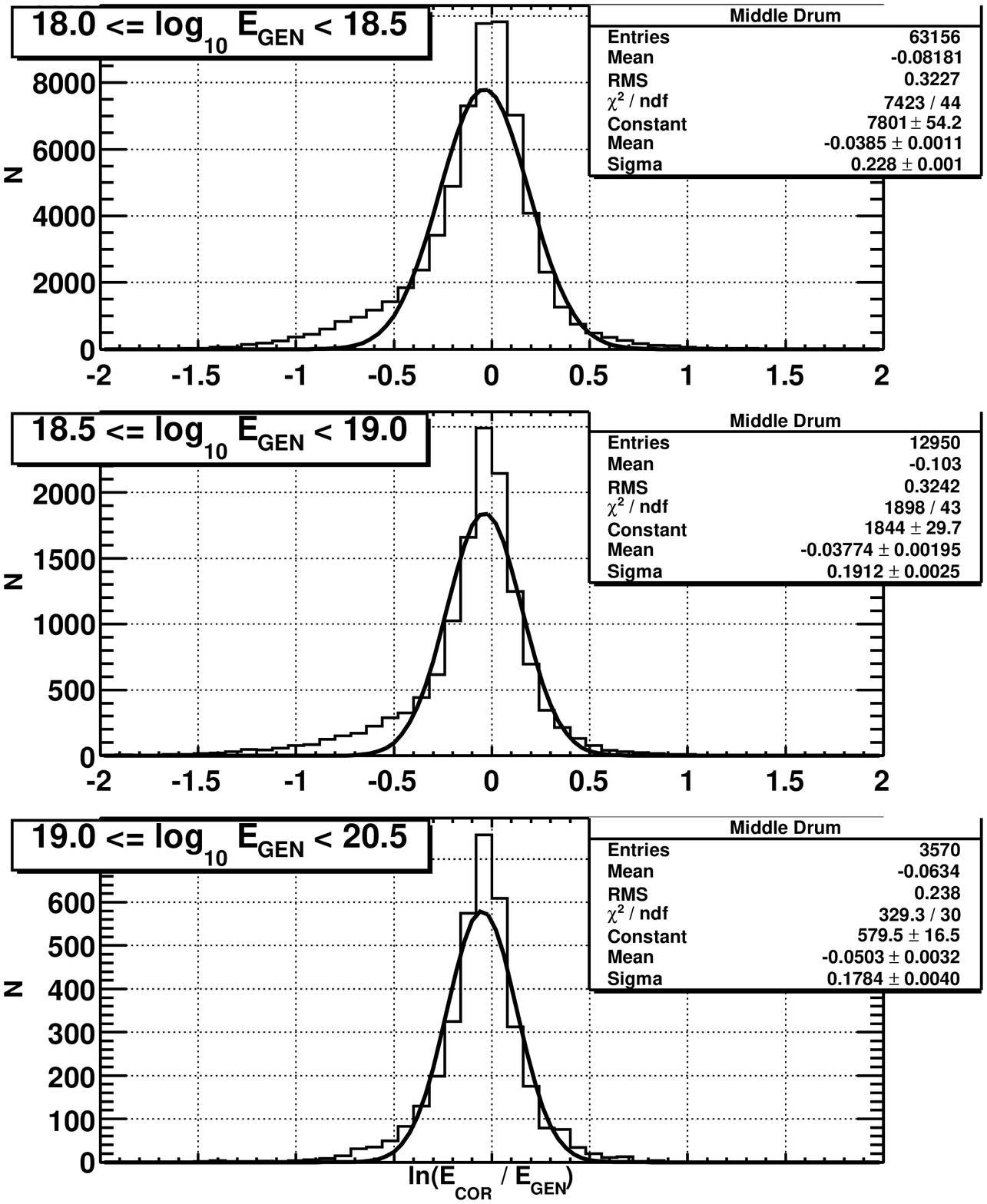}
  \caption{Resolution in the measurement of the energy of Monte Carlo
    simulated events shown in three energy ranges: $10^{18.0-18.5}$ eV, $10^{18.5-19.0}$ eV, and $> 10^{19.0}$ eV. The gaussian fit is
    used to determine the detector bias and resolution.} 
  \label{fig:Res_En}
\end{figure}

Resolution plots indicate how well the detector simulation and
reconstruction programs perform by comparing reconstructed values to
generated values in Monte Carlo simulated events. The
three primary parameters that show the quality of the reconstruction
are the impact parameter ($R_{P}$) and the in-plane
angle ($\psi$) obtained from the geometrical reconstruction, and the
energy, obtained from the profile reconstruction. These are determined
for the same three energy ranges as the data-Monte Carlo comparisons
to show trends in the reconstruction. With increasing reconstructed
energy, the geometrical parameters show a trend of improving resolution (see Figures \ref{fig:Res_Rp} and
\ref{fig:Res_Psi}). For all energy
ranges, the energy resolution is on the order of 20\% (see Figure \ref{fig:Res_En}).

\section{The Energy Spectrum}
\label{Spectrum}

\begin{figure}[htp]
  \centering
  \includegraphics[width=\textwidth]{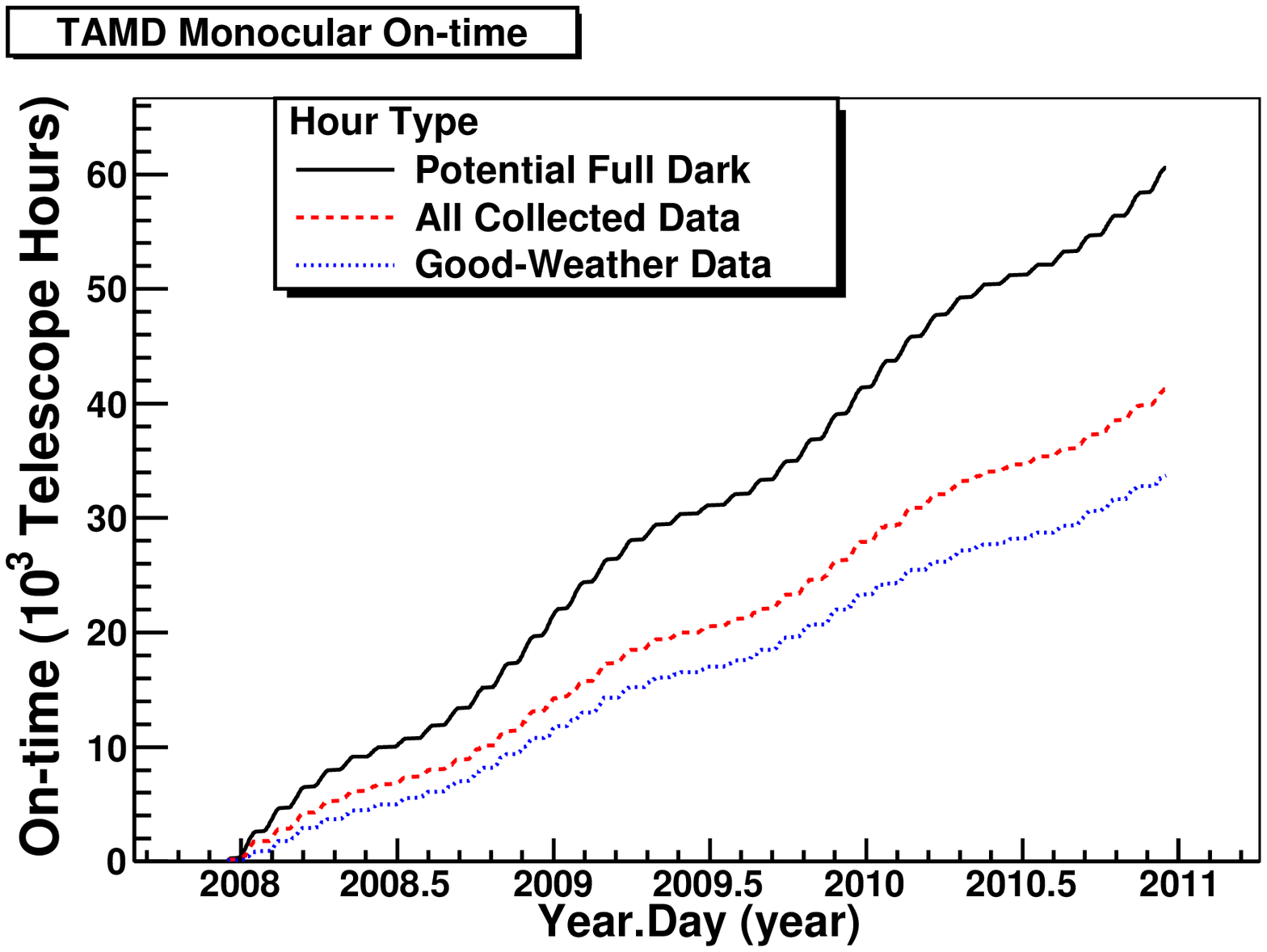}
  \caption{Integrated data collection time as a function of elapsed
    time for the Middle Drum fluorescence telescope site.} 
  \label{fig:TAMD_Ontime}
\end{figure}

The measured energy flux spectrum includes data
collected using the Middle Drum fluorescence telescope station between December 16, 2007 and December 16, 2010 (see Figure
\ref{fig:TAMD_Ontime}). The spectrum only includes data collected on
clear, moonless nights with minimal cloud cover in the view of the
detector for reliable reconstruction. This amounts to $\sim 2400$
site-hours of data collection, corresponding to a 9\% duty
cycle. Multiplying this on-time with the aperture determines the Middle Drum exposure
to be $\sim1/3$ that of the final HiRes-1 exposure.

\begin{figure}[htp]
  \centering
  \includegraphics[width=\textwidth]{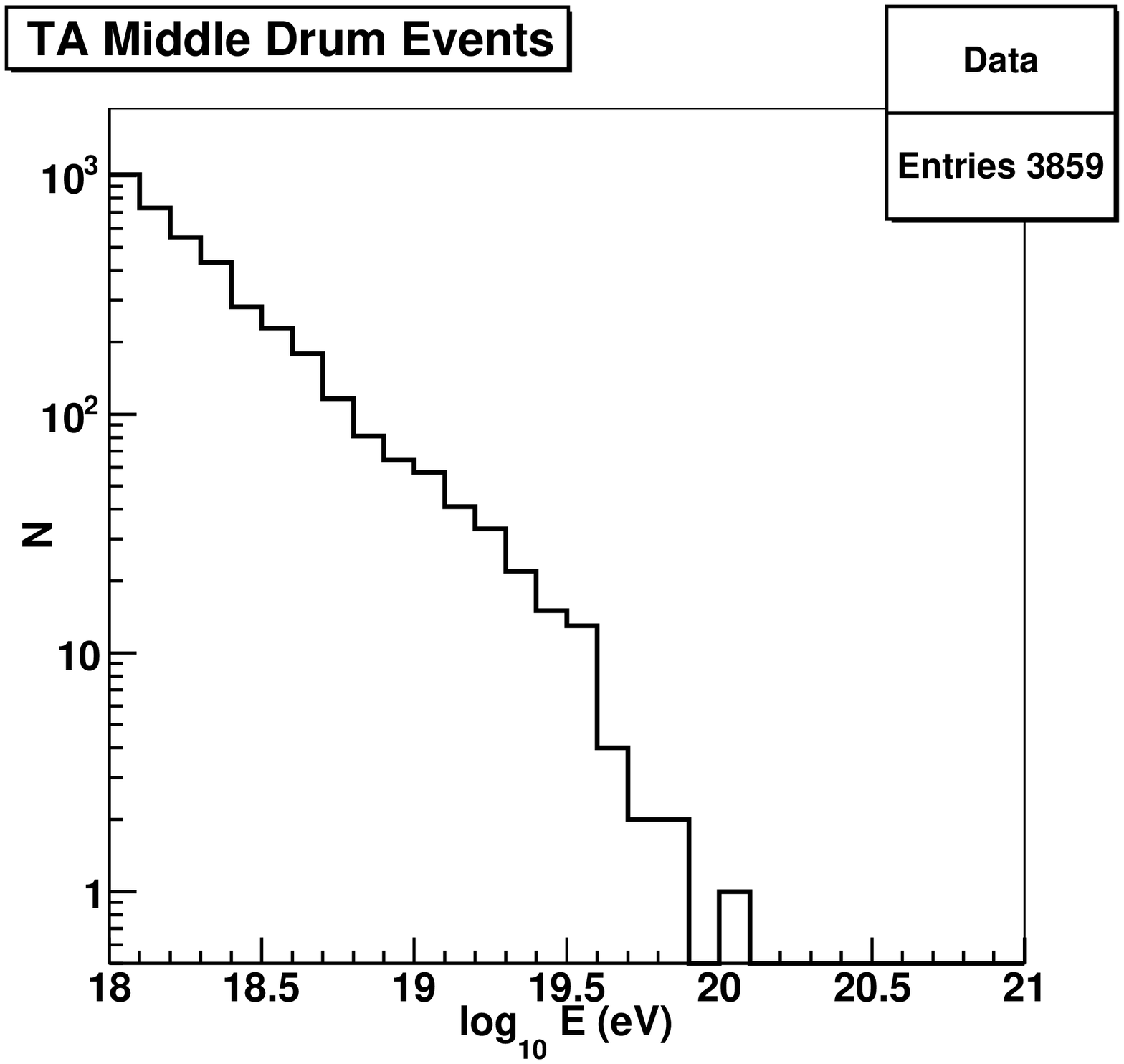}
  \caption{The raw energy distribution of events observed by the Middle Drum
    detector over the first three years of collection. The events are
    shown distributed in tenth-decade energy bins between $10^{18.0}$
    eV and $10^{20.0}$ eV} 
  \label{fig:TAMD_NEvents}
\end{figure}

After three years of collecting data, 3859 events were observed. For each energy bin in which Middle Drum has
observed events, the average number of events is $\sim32.5\%$ that observed by
HiRes-1. This is consistent with the Middle Drum 
exposure calculation. As was mentioned previously, an inverse-Monte Carlo technique is used
in order to determine the energy of the shower. The Monte Carlo shower library,
parameterized by the Gaisser-Hillas equation (see Equation
\ref{eq:Gaisser-Hillas_Eq}), is sampled for similar $X_{max}$
values and projected along the calculated geometry. The signal for
each slant-depth bin of the simulated shower is then compared to the
observed shower and chi-square values are calculated for a series of $psi$ angles. The minimum
chi-square value for a combined geometry and profile is determined to be the
best-fit reconstruction. The showers are then
distributed into tenth-decade energy bins from $10^{18.0} - 10^{21.0}$
eV. The raw energy distribution for the data is shown in Figure
\ref{fig:TAMD_NEvents}.

\begin{figure}[htp]
  \centering
  \includegraphics[width=\textwidth]{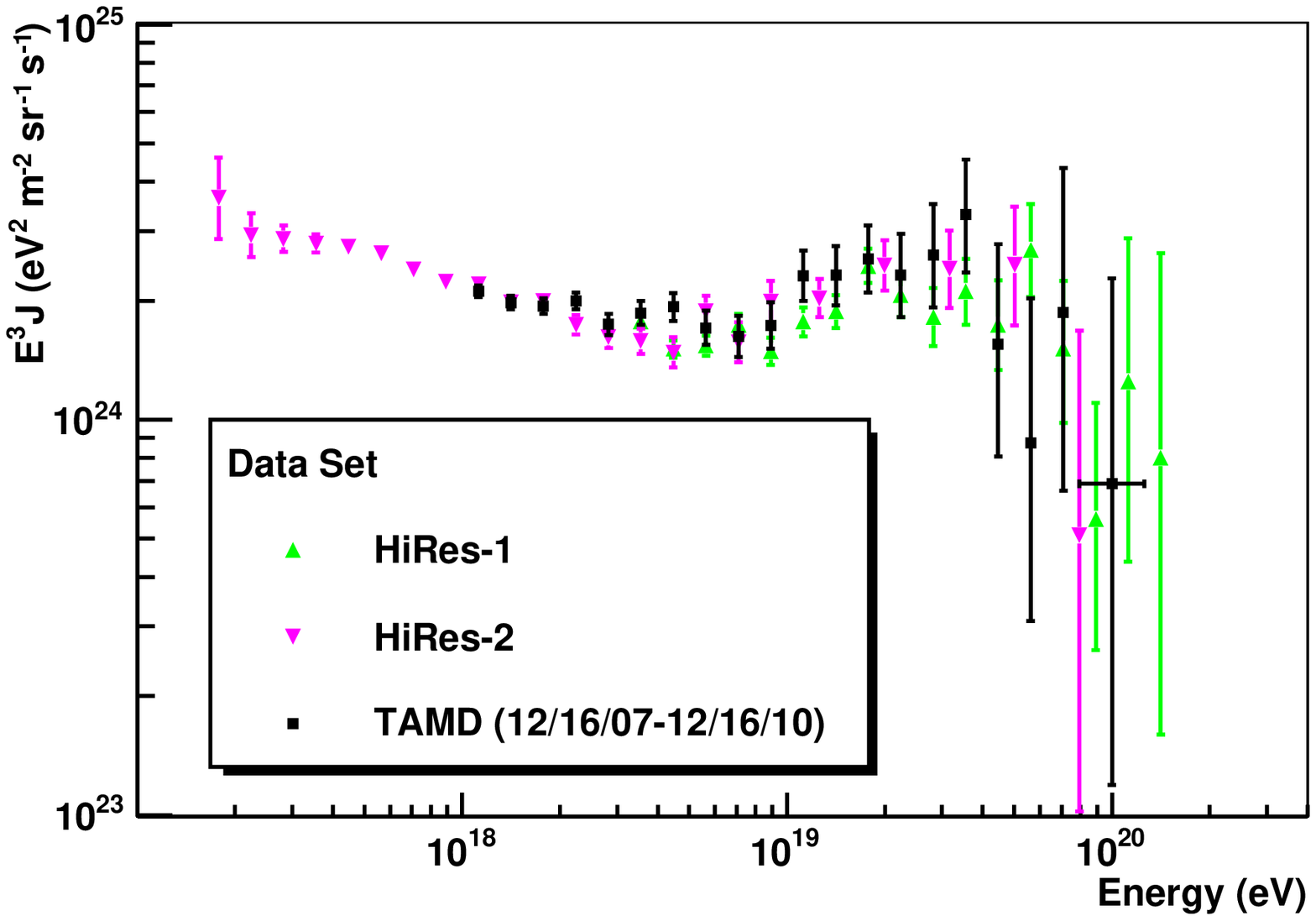}
  \caption{The energy spectra multiplied by $E^{3}$. The spectrum as
    determined from the Middle Drum data is shown by the black
    boxes. The spectra of HiRes-1 (upward, green triangles) and HiRes-2
    (downward, magenta triangles) are shown for comparison. The three
    spectra are in excellent agreement in both normalization and shape.} 
  \label{fig:TAMD_Spect}
\end{figure}

\begin{figure}[htp]
  \centering
  \includegraphics[width=\textwidth]{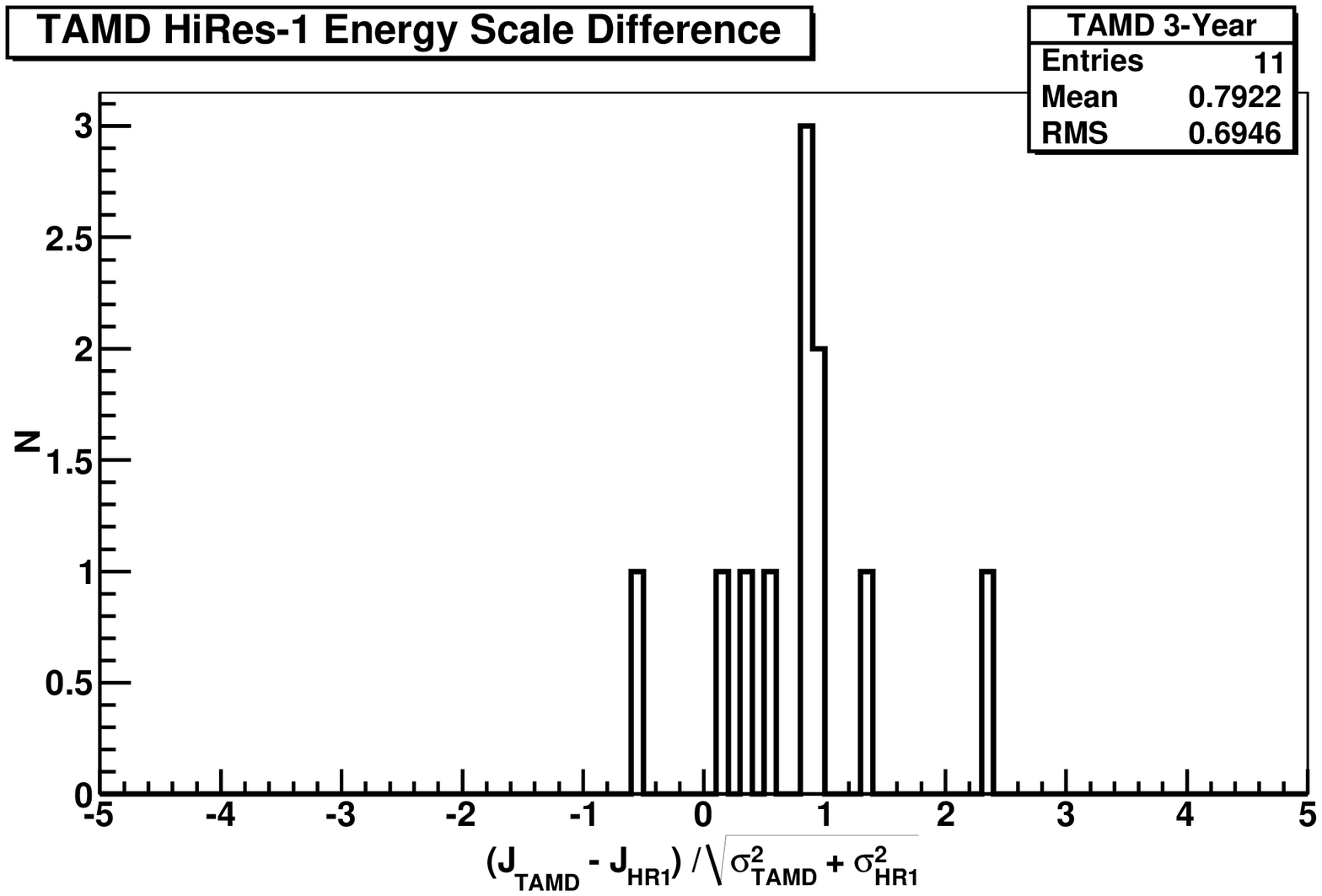}
  \caption{The bin-by-bin energy flux measurement difference between Middle Drum
    and HiRes-1. Only energy bins between $10^{18.55}$ and $10^{19.55}$ eV were
    compared since these are the only bins in the Middle Drum data
    where there are more than seven observed events within the HiRes-1
    energy range.} 
  \label{fig:TAMD_HR1_Diff}
\end{figure}

The flux is calculated by combining the number of events and the exposure per energy bin using the equation
\begin{equation}
  J(E) = \frac{n(E)}{A\Omega(E) \times \Delta t_{on} \times \Delta E},
  \label{eq:Flux_calculation}
\end{equation}
\noindent where $n(E)$ is the number of events in a given energy bin,
$E$; $\Delta E$ is the width of the energy bin; $A\Omega(E)$ is the
energy-dependent aperture calculated from Equations \ref{eq:Aper_Zero}
and \ref{eq:Aperture_calc}; and $\Delta t_{on}$ is the on-time of the
detector. This flux is often multiplied by the cube of the energy to flatten
the spectrum in order to more clearly show the subtle features of the flux of
these particles. Figure \ref{fig:TAMD_Spect} shows the spectrum as
determined from the Middle Drum data overlaid with that from the
two HiRes detectors' monocular reconstructions
\cite{HiRes_GZK_paper}. These are in excellent agreement in both
normalization and shape. A consistency between the spectra measured by
the Middle Drum detector and the HiRes-1 detector is
determined by:
\begin{equation}
  \Delta J = \frac{ \left( J_{TAMD} -
        J_{HiRes-1} \right)}{\sqrt{\sigma_{TAMD}^{2} +
        \sigma_{HiRes-1}^{2}}}
    \label{eq:TAMD_HR1_Diff}
\end{equation}
where $J_{TAMD}$ and $J_{HiRes-1}$ are the measured flux and
$\sigma_{TAMD}$ and $\sigma_{HiRes-1}$ are the statistical
uncertainties observed by the Middle Drum and HiRes-1 detectors,
respectively (see Figure \ref{fig:TAMD_HR1_Diff}). This calculation
only included those energy bins with at least seven events observed 
by the Middle Drum telescopes within the HiRes-1 spectral range. This
difference shows that the flux measured by Middle Drum is within
$0.8\sigma$ of HiRes-1. They are consistent with the same flux level.

Further, the Middle Drum spectrum can be quantitatively compared to the HiRes-1 spectrum
by determining the $\chi^{2}$ between the flux measurements on a
bin-by-bin basis. The $\chi^{2}$ value is calculated by summing over
the square of each $\Delta J$ given in Equation \ref{eq:TAMD_HR1_Diff}.
The result is a $\chi^{2}/N.D.F. = 12.20 / 10$ for all of the overlapping bins and
$\chi^{2}/N.D.F. = 4.47 / 5$ for bins $\geq 10^{19.0}$ eV. Figure
\ref{fig:TAMD_Spect} shows that the $\chi^{2}$ is dominated
by the difference in the measured flux in the $10^{18.65}$ eV energy
bin. 

Since Middle Drum measured the spectrum with the same equipment
and calibration techniques and obtained the same result as HiRes-1,
the HiRes-1 energy scale is thus transferred to Middle Drum. Had the
energy scale changed, the rapidly falling $E^{-3}$ spectrum would have
shifted by twice that increment.

\section{Middle Drum Hybrid Geometry Comparison}
\label{Hybrid}

The transfer of the energy scale from the HiRes-1 spectrum to the
Middle Drum spectrum creates a direct link between the HiRes and
Telescope Array experiments. The next step to completely bridge
the two experiments is to determine the energy scale between those
events observed by the Middle Drum detector to those that also
triggered the ground array. This Middle Drum monocular-hybrid
comparison will then transfer the energy scale of HiRes to
the rest of Telescope Array in future studies. 

The hybrid analysis begins by improving the geometrical
reconstruction. Time and pulse height information of the triggered
ground array scintillator detectors (SDs) are used to improve the
time-versus-angle fit \cite{Monica_ICRC2011}. This
is performed by calculating the SD core 
using the modified Linsley shower-shape \cite{SD_Fitting} to obtain a
lateral distribution function (LDF) which is then used to constrain
the monocular time-versus-angle fit performed for the FD.

After the improved shower geometry is determined, the profile fit is performed
using the inverse-Monte Carlo technique presented in
this paper, however, in this fit, the geometry determined above was not adjusted to
scan for a better profile fit. The hybrid data selection cuts use a combination of
Middle Drum and SD information. Events are retained if:
\begin{enumerate}
\item the profile fit reconstructs well, as is determined in the
  monocular reconstruction;
\item the geometry fit has a $\chi^{2}/NDF < 7$;
\item the zenith angle is $< 56^{\circ}$, providing a
  well-reconstructed SD core impact location of simulated showers
  thrown up to $60^{\circ}$;
\item the SD calculated core must be within 500 meters of the SD
  boundary, so there is no bias in the LDF reconstruction;
\item the SD calculated core must be within 600 meters of the
  shower-detector plane, so the shower track remains consistent
  between the two detectors;
\item the angular track length is $> 7.9^{\circ}$, to provide a
  reliable profile fit; and
\item $X_{max}$ is observed by Middle Drum, for reliable
  composition studies.
\end{enumerate}

\begin{figure}[htp]
  \centerline{
    \subfigure[$10^{18.0}-10^{18.1}$ eV]{
      \includegraphics[width=0.5\textwidth]{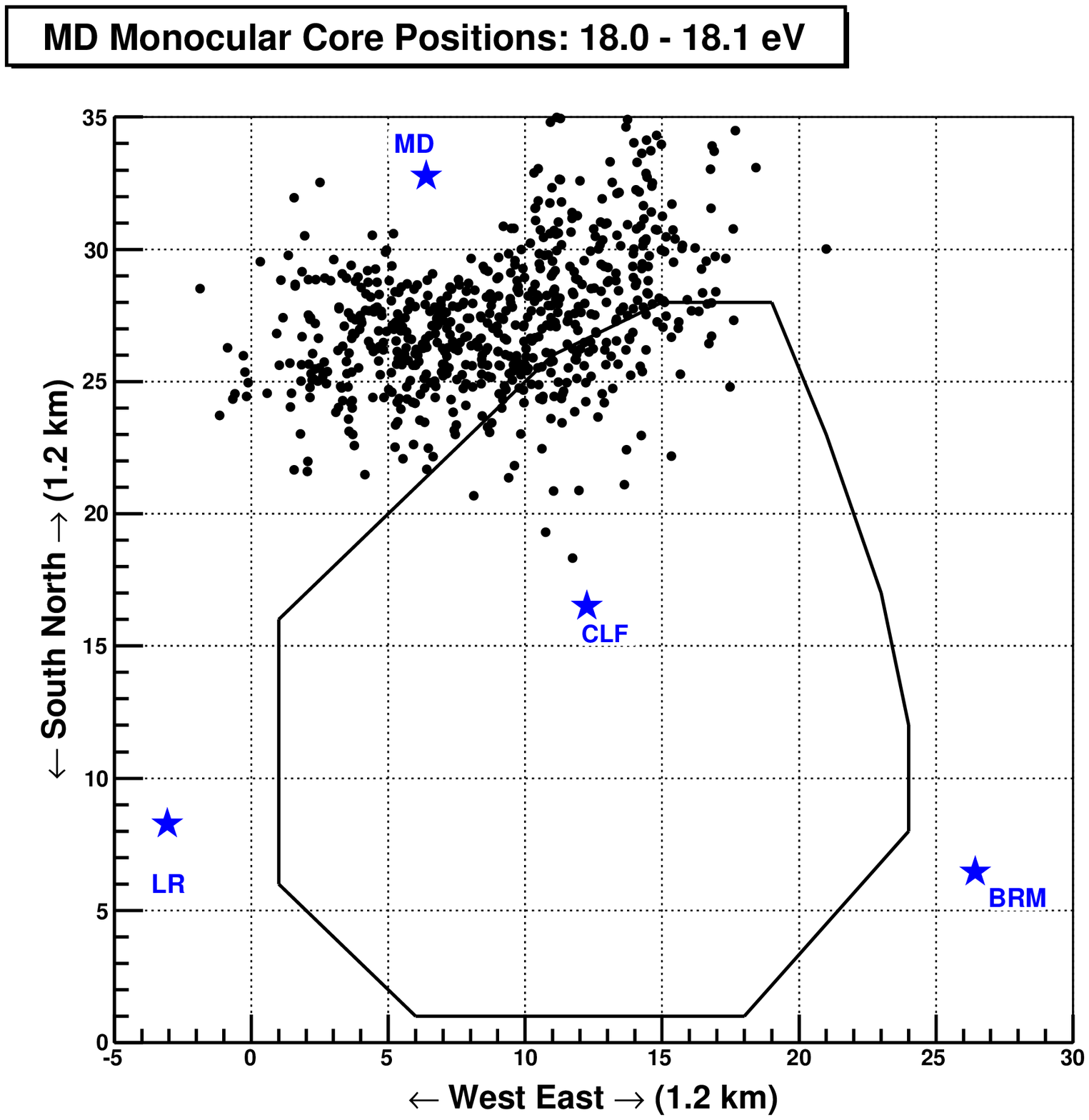}
      \label{fig:TAMD_Hybrid_180-181}
    }
    \subfigure[$10^{18.1}-10^{18.2}$ eV]{
      \includegraphics[width=0.5\textwidth]{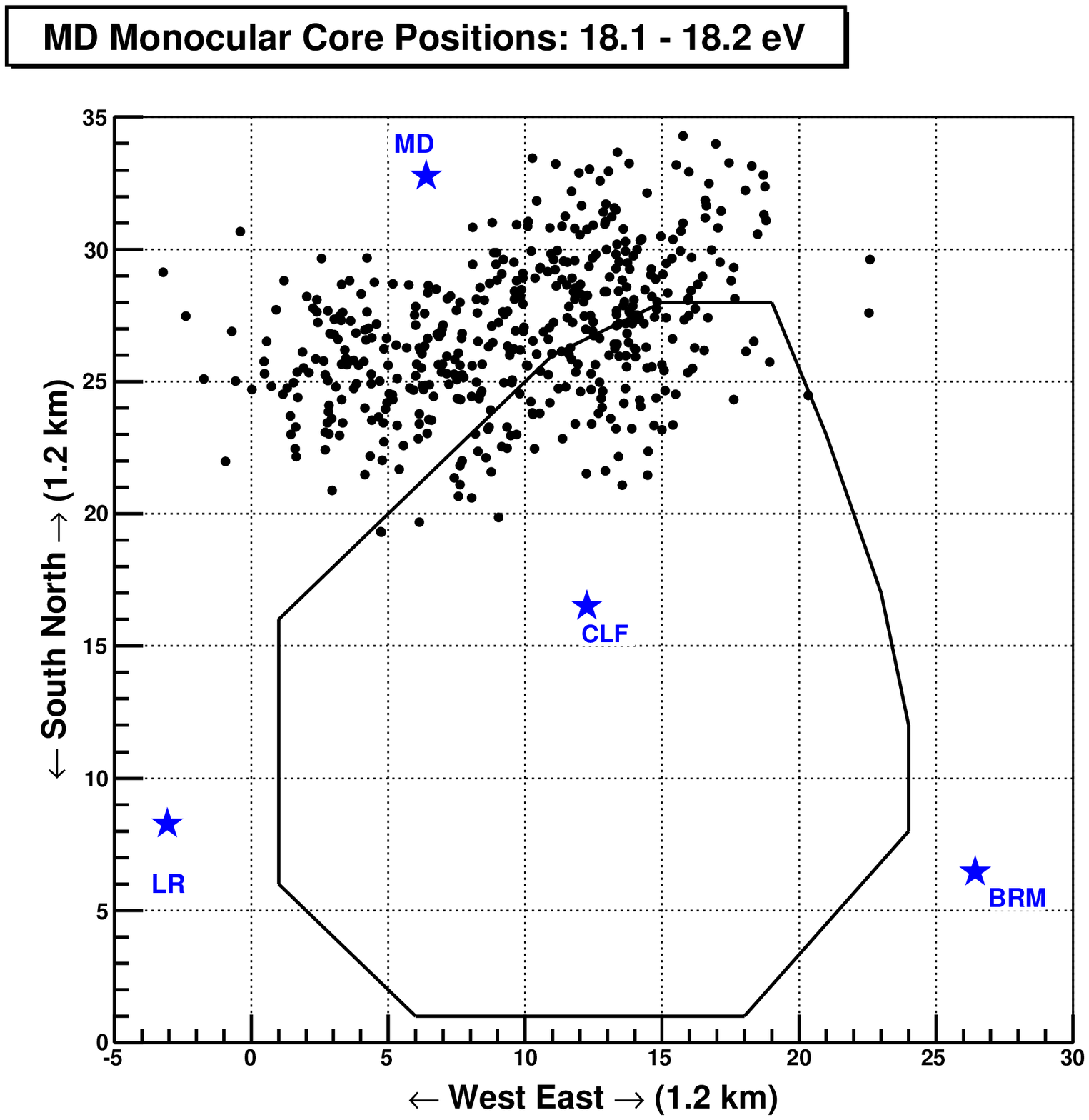}
      \label{fig:TAMD_Hybrid_181-182}
    }
  } 
  \centerline{
    \subfigure[$10^{18.2}-10^{18.3}$ eV]{
      \includegraphics[width=0.5\textwidth]{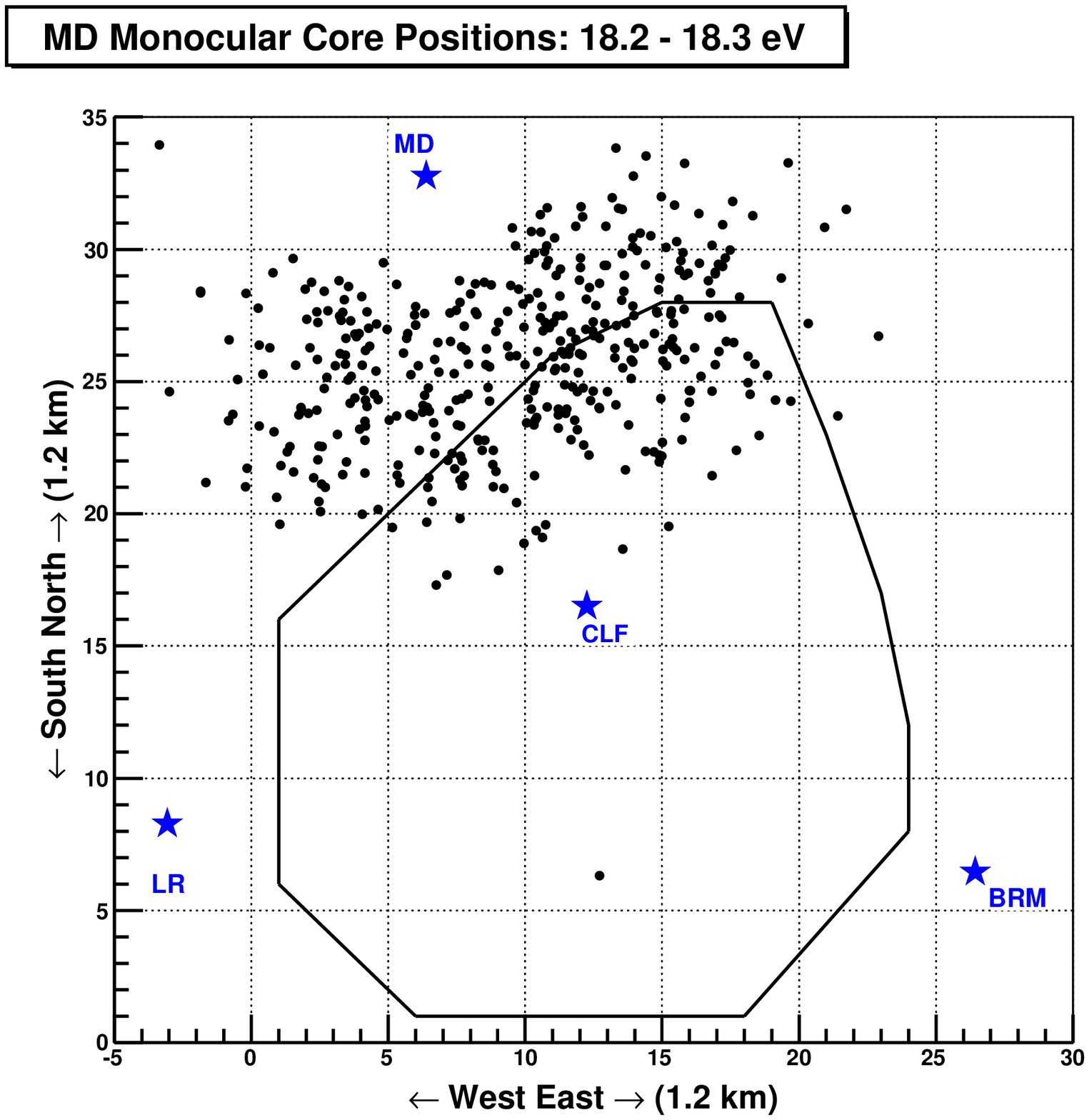}
      \label{fig:TAMD_Hybrid_182-183}
    }
    \subfigure[$10^{18.3}-10^{18.4}$ eV]{
      \includegraphics[width=0.5\textwidth]{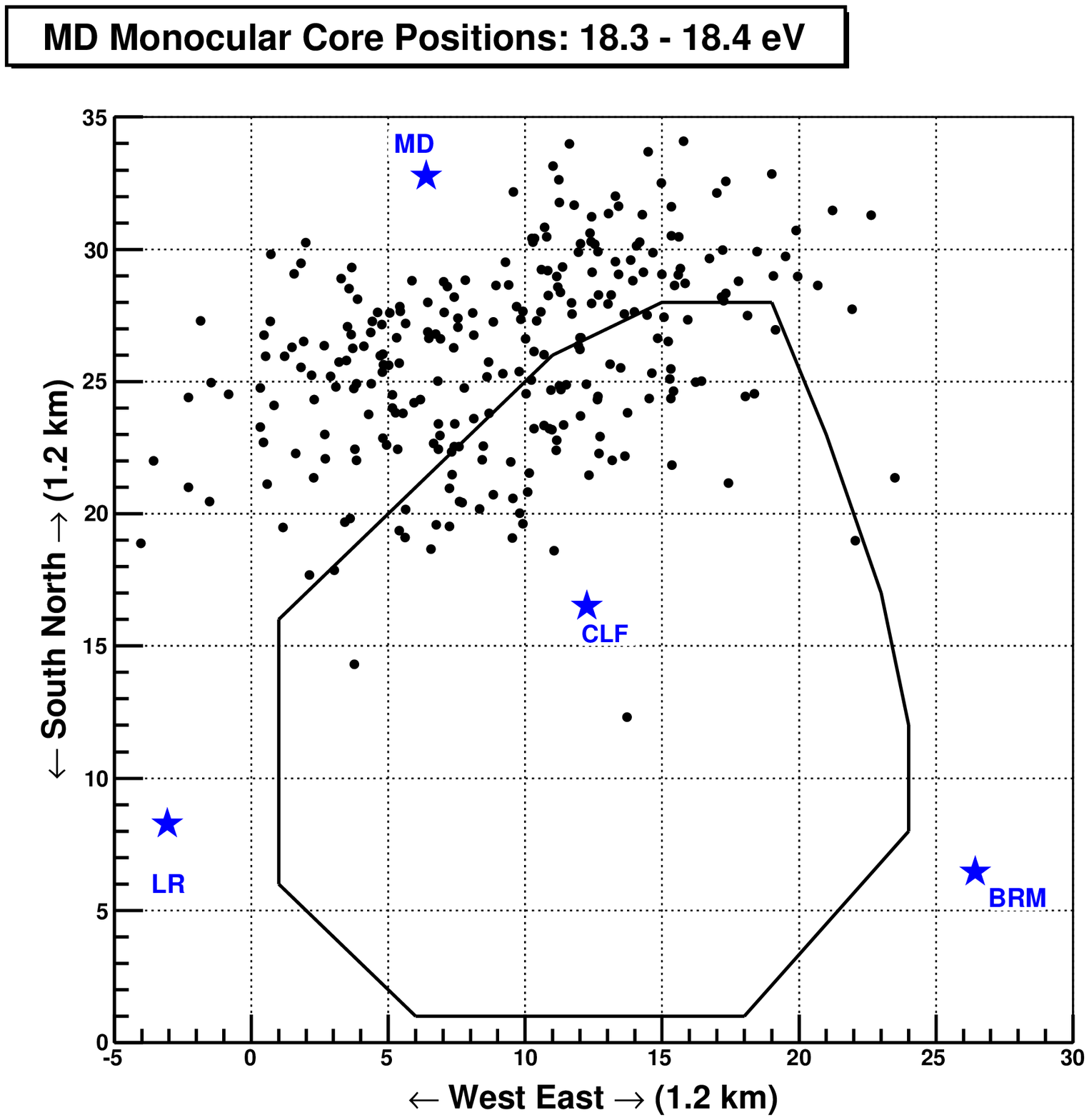}
      \label{fig:TAMD_Hybrid_183-184}
    }
  }
  \caption{The core positions of the Middle Drum events observed and
    reconstructed in monocular mode (indicated by the points) for tenth-decade energy bins between $10^{18.0}$ and
    $10^{18.4}$ eV. The locations of the
    fluorescence stations (BRM, LR, and MD), as well as the Central
    Laser Facility (CLF), are indicated by stars. The perimeter of the
    scintillator Surface Detector (SD) array is indicated with lines. At
    these lowest energies, virtually all core locations are outside of the
    SD array.}
  \label{fig:TAMD_Hybrid_185Limit_Low}
\end{figure}

\begin{figure}[htp]
  \centerline{
    \subfigure[$10^{18.4}-10^{18.5}$ eV]{
      \includegraphics[width=0.5\textwidth]{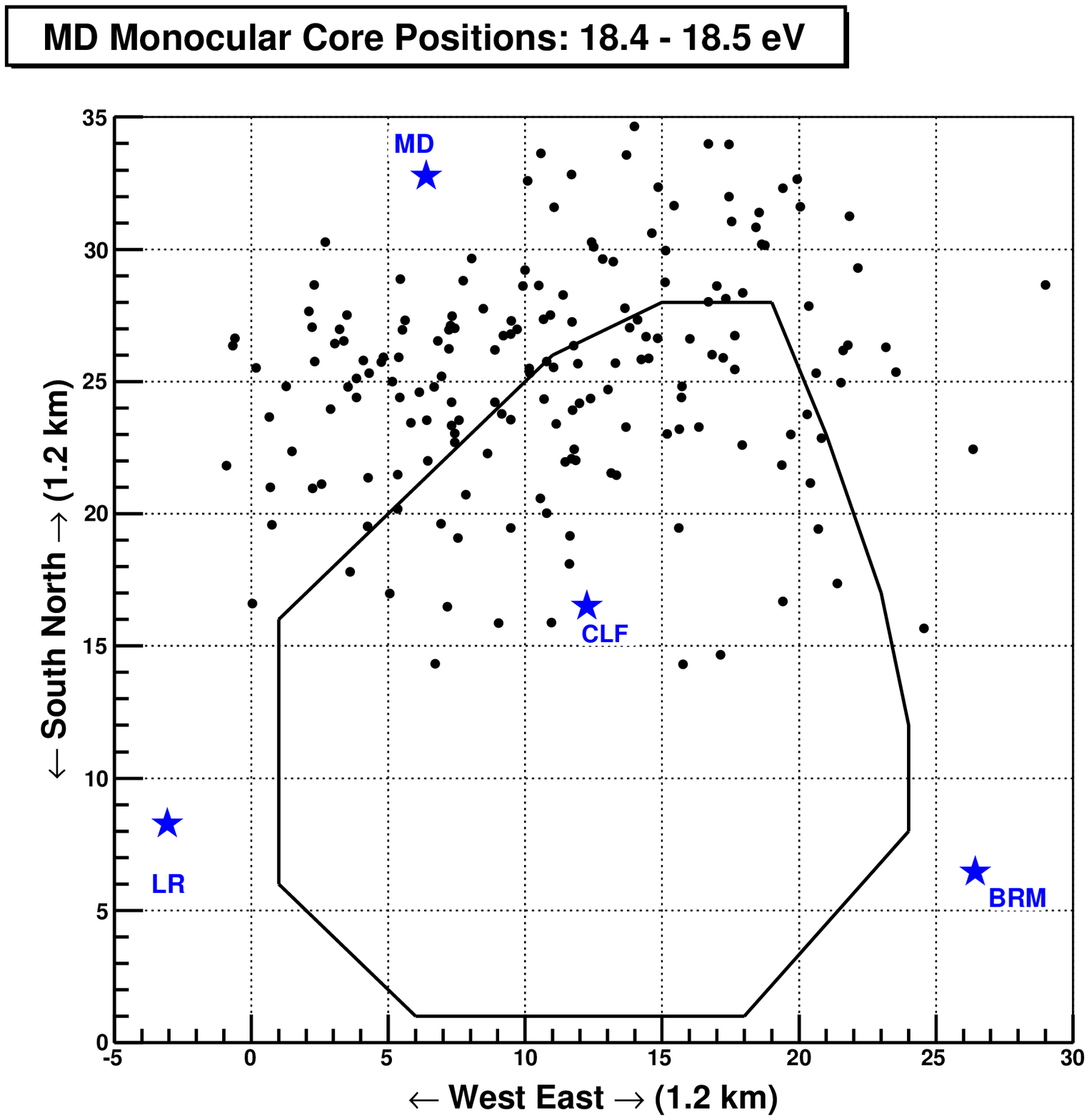}
      \label{fig:TAMD_Hybrid_184-185}
    }
    \subfigure[$10^{18.5}-10^{18.6}$ eV]{
      \includegraphics[width=0.5\textwidth]{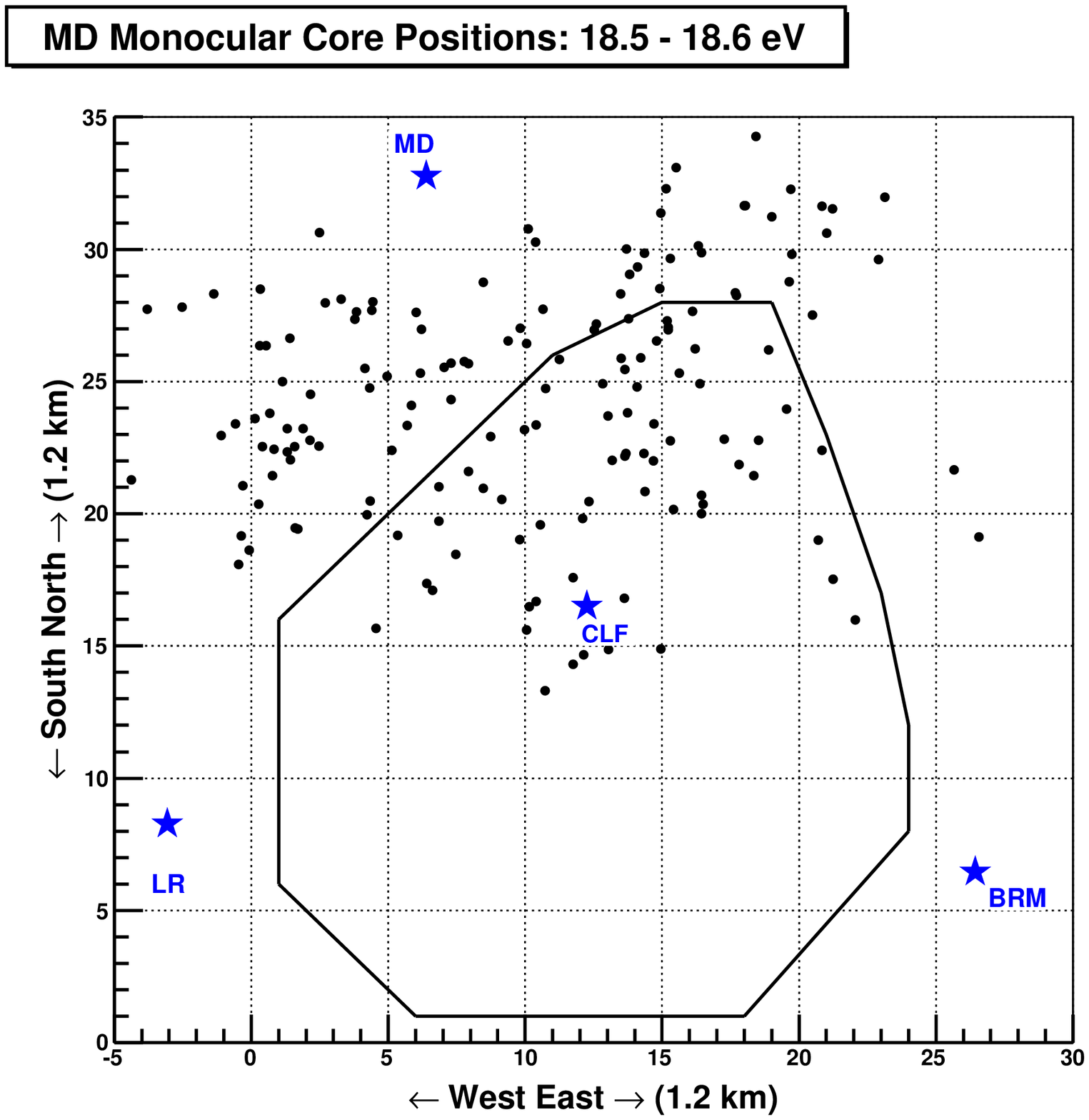}
      \label{fig:TAMD_Hybrid_185-186}
    }
  }
  \caption{The core positions of the Middle Drum events observed and
    reconstructed in monocular mode (indicated by the points) for tenth-decade energy bins between $10^{18.4}$ and
    $10^{18.6}$ eV. Compared to those events shown in Figure
    \ref{fig:TAMD_Hybrid_185Limit_Low}, as the event energy increases,
    an increasing percentage of cores are observed within the SD array.}
  \label{fig:TAMD_Hybrid_185Limit_High}
\end{figure}

Since the boundary of the ground array begins $\sim7$ km from the
Middle Drum site, most of the monocular events with
energy less than $10^{18.4}$ eV fall
outside of the ground array (see Figures \ref{fig:TAMD_Hybrid_180-181}
through \ref{fig:TAMD_Hybrid_183-184}). Above this energy, roughly
half of the events observed monocularly have core positions within the
boundary of the ground array (see Figures
\ref{fig:TAMD_Hybrid_184-185} and
\ref{fig:TAMD_Hybrid_185-186}).

\begin{figure}[htp]
  \centering
  \includegraphics[width=\textwidth]{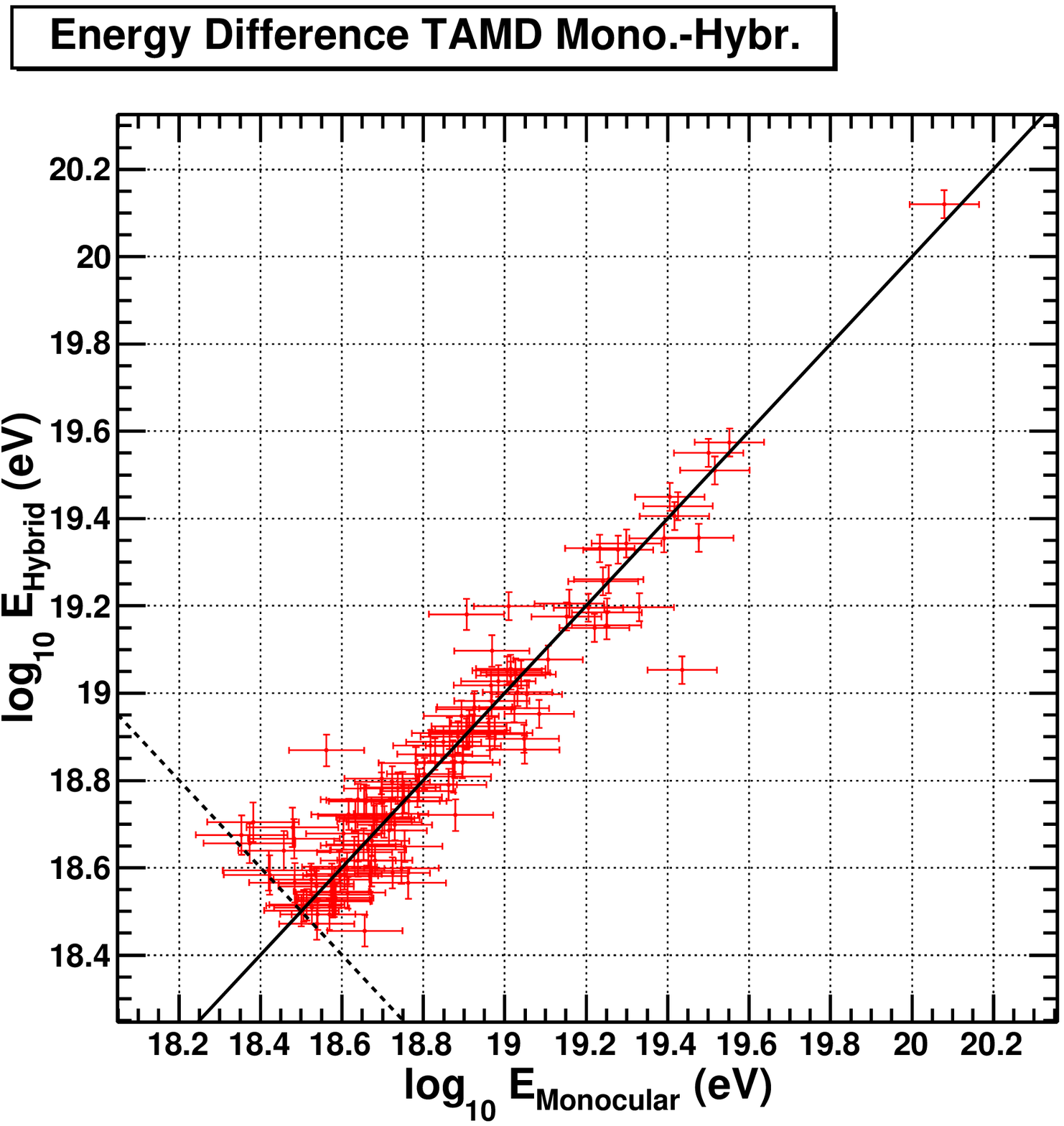}
  \caption{Comparison of event energy for cosmic ray events
    reconstructed in monocular mode using data collected from the
    Middle Drum fluorescence telescope station (abcissa) versus the
    energy when reconstructed in hybrid mode (ordinate) incorporating
    information from the scintillator surface array. The solid line
    drawn indicates where the two measurements would be equal. There
    is excellent agreement between the measurements.} 
  \label{fig:PCGF_Hybrid_Energy}
\end{figure}

\begin{figure}[htp]
  \centering
  \includegraphics[width=\textwidth]{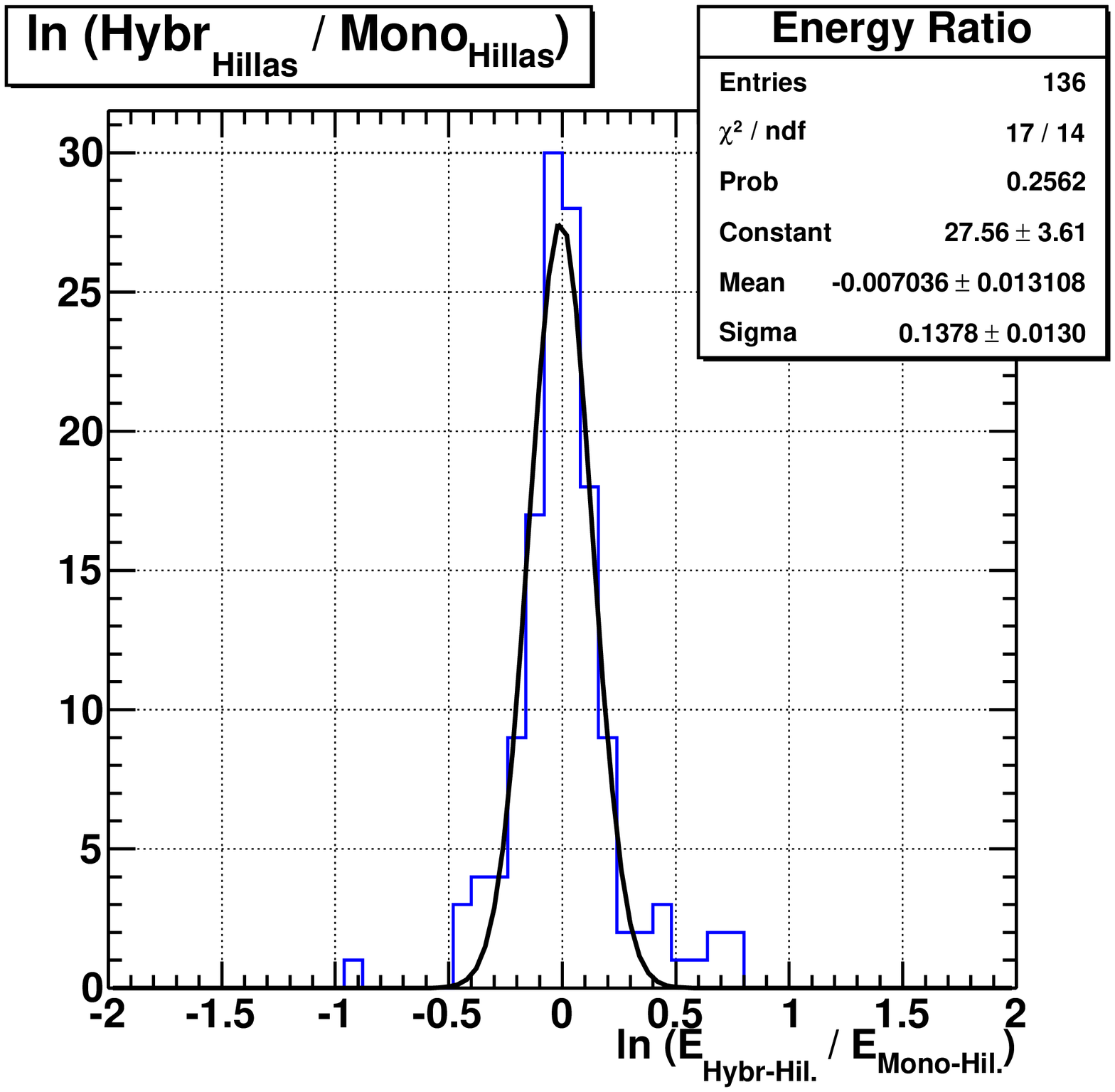}
  \caption{Spread of difference in event energy for cosmic ray events
    observed by the Middle Drum fluorescence telescope station
    reconstructed in hybrid mode incorporating information from the scintillator surface array
    to that reconstructed in monocular mode. The gaussian mean shows
    the overall difference between these reconstruction techniques,
    with an uncertainty indicated by the standard deviation. There
    is excellent agreement between the measurements.} 
  \label{fig:PCGF_Hybrid_Scale}
\end{figure}

Figure \ref{fig:PCGF_Hybrid_Energy} shows a comparison of the energy for events
reconstructed from the Middle Drum data in monocular mode to the same
events reconstructed in Middle Drum-hybrid mode. Only events retained in both the monocular and hybrid analyses were compared. For those events with
$\sqrt{E_{mono} \times E_{hybr}} > 10^{18.5}$ eV, the monocular and
hybrid energies are in good agreement
(see Figure \ref{fig:PCGF_Hybrid_Scale}). 
This provides a direct link between the events observed by
Middle Drum in monocular mode to those events that also trigger the
ground array. A direct comparison is thus made between
HiRes and all of the Telescope Array detectors. 

\section{Conclusions}
\label{Conc}

The Telescope Array's Middle Drum observatory uses
refurbished telescopes from the High Resolution Fly's Eye experiment.
A spectral measurement was made using the first three years of the
Middle Drum data collection. Both the data and simulated events were analyzed monocularly using
the profile-constrained geometry reconstruction technique that was
developed for the HiRes-1 data. The energy and geometrical resolutions of the Monte Carlo
simulations show good agreement between what was generated and what
was reconstructed and the data-Monte Carlo comparisons are in excellent
agreement between simulated and real extensive air showers. The calculated Middle Drum
energy spectrum is shown to be in excellent agreement with the spectra
produced by the HiRes-1 monocular analysis with the difference between
them less than the energy resolution of the Middle Drum
reconstruction. The HiRes energy scale can
now be transferred to the entire Telescope Array for further comparisons now under way.

\section{Acknowledgements}
The Telescope Array experiment is supported 
by the Japan Society for the Promotion of Science through
Grants-in-Aid for Scientific Research on Specially Promoted Research (21000002) 
``Extreme Phenomena in the Universe Explored by Highest Energy Cosmic Rays'', 
and the Inter-University Research Program of the Institute for Cosmic Ray 
Research;
by the U.S. National Science Foundation awards PHY-0307098, 
PHY-0601915, PHY-0703893, PHY-0758342, and PHY-0848320 (Utah) and 
PHY-0649681 (Rutgers); 
by the National Research Foundation of Korea 
(2006-0050031, 2007-0056005, 2007-0093860, 2010-0011378, 2010-0028071, R32-10130);
by the Russian Academy of Sciences, RFBR
grants 10-02-01406a and 11-02-01528a (INR),
IISN project No. 4.4509.10 and 
Belgian Science Policy under IUAP VI/11 (ULB).
The foundations of Dr. Ezekiel R. and Edna Wattis Dumke,
Willard L. Eccles and the George S. and Dolores Dore Eccles
all helped with generous donations. 
The State of Utah supported the project through its Economic Development
Board, and the University of Utah through the 
Office of the Vice President for Research. 
The experimental site became available through the cooperation of the 
Utah School and Institutional Trust Lands Administration (SITLA), 
U.S.~Bureau of Land Management and the U.S.~Air Force. 
We also wish to thank the people and the officials of Millard County,
Utah, for their steadfast and warm support. 
We gratefully acknowledge the contributions from the technical staffs of our
home institutions and the University of Utah Center for High Performance Computing
(CHPC).

\end{document}